# Effective properties of composites with open boundary conditions

Guo-Qing Gu [1], En-Bo Wei [2,a]

[1] *School of Computer Science and Technology, East China Normal University, Shanghai 20062, People's Republic of China*

[2] *Institute of Oceanology, Chinese Academy of Sciences and Qingdao National Laboratory for Marine Science and Technology, Qingdao 266071, People's Republic of China*

[a] Author to whom correspondence should be addressed: ebwei@qdio.ac.cn

**Abstract** A primitive problem of predicting the effective properties of composites is open boundary conditions. In this paper, Eshelby's transformation field method is developed to solve the open boundary problem of two-phase composites having arbitrary inclusion geometry. The transformation fields are introduced in the composite system to cope with the complicated interface boundary-value problem of the composite. Furthermore, the open boundary problem is solved by Hermite polynomial, which is used to express the transformation fields and the perturbation fields. As an example, the formulas of Eshelby's method are derived for calculating the effective dielectric property of two-dimensional isotropic dielectric composites having open boundary conditions. The validity of the method is verified by comparing the effective responses estimated by the method with the exact solutions of dilute limit. The results show that the method is valid to solve the open boundary problem of composites having complex geometric inclusions.

## 1. Introduction

Many of research methods were proposed to deal with the effective properties of composites, such as self-consistent model [1-4], variational approach [5,6], statistical method and exact solutions [7,8]. In various composites, there is one kind of composites of the inclusions embedded in an infinite matrix. For the composites having the regular geometric inclusions embedded in the infinite matrix, some analytical solutions are derived for the problems of electric, elastic, piezoelectric and thermal composites, such as cylindrical, ellipsoidal and spherical composites [9-17]. The open boundary problem of the composites is more primitive and very difficult to be analytically solved for the composites having complex structure inclusions because of the complex interactions between the inclusion and matrix materials. To investigate the open boundary problem of the composites, it is possible to develop Eshelby's transformation field method to solve the effective responses of the composites having arbitrary inclusion shapes [18-24].

Eshelby's method had been proposed to calculate the elastic tensors of whole ellipsoidal composite regions [18,19], where the elastic tensor can be linearly expressed by the uniform transformation strain (i.e. stress-free strain) introduced by an ellipsoid inclusion through the Eshelby's uniform tensor, which contains the geometric parameters and physical properties of ellipsoidal inclusions. Considering the non-uniform transformation strain tensor, Nemat-Nasser et al. [20,21] had developed Eshelby's method to estimate the overall stress and the effective elastic moduli of periodic composites having arbitrary inclusion geometry. Subsequently, Eshelby's method was first extended by Gu et al.[25-29] to predict the effective electric conductivity, viscosity, piezoelectric properties and dispersion relation of photonic band structure of periodic composites. Jiang et al.[17] obtained the closed form solutions of the electro-elastic Eshelby's tensors of a piezoelectric ellipsoidal inclusion in an infinite non-piezoelectric matrix via the Green's function technique. Moreover, Zhou [30] calculated the effective elastic responses of periodic composites having multiple inhomogeneous inclusions of arbitrary shapes. For the graded periodic composites, the method was used to obtain the effective dielectric constants of periodic composites with arbitrary gradient and geometric inclusions [31,32]. Recently, Eshelby's method is developed to discuss the effective responses of unidirectional periodic composites having an open boundary condition [33]. However, although Eshelby's transformation field method has successfully solved some boundary problems of periodic composites proposed by Rayleigh [34], the open boundary problem is not solved by Eshelby's method for the composites of arbitrary structure inclusions embedded in the infinite matrix. Along this line, we investigate Eshelby's transformation field method to solve the open boundary problem of the composites having arbitrary inclusion shapes. Applied an external field to the composite, the open boundary conditions are always used since the perturbation fields induced by inclusions tend to zero at infinite point, which is far from the inclusion region. The fields at the infinite point of the matrix regions regarded as the external applied field is an intrinsic open boundary condition in physics.

In this paper, Eshelby's transformation field method is proposed to estimate the effective properties of the isotropic dielectric composites having the open boundary conditions. The transformation electrical fields in the composite regions are introduced to set up a unified constitutive relation. To match the open boundary conditions, the perturbation potential and the transformation electrical fields are expressed in terms of Hermite polynomials. Furthermore, the transformation electric fields are obtained by solving a set of algebraic equations with respect to the coefficients of the transformation fields. The formulas of the effective dielectric responses are expressed by the transformation fields. Finally, the validity of the method is verified by comparing the effective responses estimated by the method with the exact solutions of dilute limit.

## 2. Formulation

### *2.1. Transformation field and its governing equation*

Consider two-phase isotropic dielectric composites, where the inclusions of arbitrary shapes are embedded in an infinite matrix. For two-dimensional problem of the composites, we assume that the isotropic constitutive relations of both matrix and inclusion are linear in Cartesian coordinates $(x, y)$,

$$D_k^a = e^a E_k^a(x, y), \qquad (1)$$

where $k = x, y$. $a = h, i$ denote quantities of the matrix and inclusions, respectively. $D$, $E$ and $e$ are the electrical displacement, electrical field and dielectric constant, respectively.

The electrostatic governing equations are $\nabla \cdot D^a = 0$ and $\nabla \times E^a = 0$. If an external uniform electric field $E_k^0$ is applied to the composites along the $k$-direction, the perturbation electric field $E_k^p$, induced by the inclusions, will occur in whole composite region. In the inclusion region $\Omega_i$ and the host region $\Omega_h$, the constitutive relations are given by $D_k^i = e^i(E_k^0 + E_k^p)$ and $D_k^h = e^h(E_k^0 + E_k^p)$, respectively.

To avoid matching the interfaces continuous conditions of the electrical displacements between inclusion and matrix, a transformation electrical field $E_k^*$ is introduced to the composite according to Eshelby's transformation field method so that the transformation electrical fields obey the following equations in the matrix and inclusion regions, respectively [18,21],

$$E_k^*(x, y) = 0 \quad \text{in} \quad \Omega_h, \tag{2}$$

$$e^i[E_k^0 + E_k^p(x, y)] = e^h[E_k^0 + E_k^p(x, y) - E_k^*(x, y)] \quad \text{in} \quad \Omega_i, \tag{3}$$

where $k = x, y$. Based on Eqs.(2) and (3), a unified constitutive relation is given in the whole composite region $\Omega = \Omega_i + \Omega_h$.

$$D_k = e^h[E_k^0 + E_k^p(x, y) - E_k^*(x, y)] \quad \text{in} \quad \Omega. \tag{4}$$

Then, the governing equation of the perturbation electrical field $E_k^p$ and the transformation electrical field $E_k^*(x, y)$ is rewritten as follows,

$$\nabla_k \{e^h[E_k^0 + E_k^p(x, y) - E_k^*(x, y)]\} = 0. \tag{5}$$

Here, note that Eq.(5) gives the relationships between the perturbation electric field and the transformation electrical field. Next, the perturbation potential is derived and expressed by the transformation electrical field from Eq. (5).

## 2.2. Solutions of perturbation potential

For the composites, in order to match the infinite boundary condition of the perturbation potential $\Phi(x, y)$ i.e. the perturbation electrical field $\vec{E}^p(x, y) = -\nabla\Phi$ and $\dot{E}^p(x, y)\big|_{x,y\to\infty} = 0$, the perturbation potential $\Phi(x, y)$ and the transformation field $E_k^*(x, y)$ are expanded in terms of Hermite polynomials,

$$\Phi(x, y) = \sum_{m=0}^{\infty}\sum_{n=0}^{\infty} \Phi_{m,n} u_m(x) v_n(y), \qquad (6)$$

$$E_x^*(x, y) = \sum_{m=0}^{\infty}\sum_{n=0}^{\infty} a_{m,n} u_m(x) v_n(y), \qquad (7)$$

$$E_y^*(x, y) = \sum_{m=0}^{\infty}\sum_{n=0}^{\infty} b_{m,n} u_m(x) v_n(y), \qquad (8)$$

where $\Phi_{m,n}$, $a_{m,n}$ and $b_{m,n}$ are unknown coefficients. The orthogonal function $u_m(x)$ is

$u_m(x) = N_m H_m(x)\exp(-\frac{1}{2}x^2)$, where $H_m(x)$ is Hermite polynomial with the coefficient, $N_m = (\frac{1}{2^m m!\sqrt{p}})^{1/2}$. $v_n(y) = N_n H_n(y)\exp(-\frac{1}{2}y^2)$. Substituting Eqs.(6), (7) and (8) into Eq.(5), we get the following Eq.(9),

$$\begin{aligned}
&\sum_{m,n=0}^{\infty} \Phi_{m,n}[\tilde{M}_m^+ u_{m+2}(x) + \tilde{M}_m^0 u_m(x) + \tilde{M}_m^- u_{m-2}(x)]v_n(y) \\
&+ \sum_{m,n=0}^{\infty} \Phi_{m,n}[\tilde{M}_n^+ v_{n+2}(y) + \tilde{M}_n^0 v_n(y) + \tilde{M}_n^- v_{n-2}(y)]u_m(x) \\
&= -\sum_{m,n=0}^{\infty} a_{m,n}[\tilde{N}_m^+ u_{m+1}(x) + \tilde{N}_m^- u_{m-1}(x)]v_n(y) \\
&- \sum_{m,n=0}^{\infty} b_{m,n}[\tilde{N}_n^+ v_{n+1}(x) + \tilde{N}_n^- v_{n-1}(x)]u_m(x)
\end{aligned} \qquad (9)$$

where $\tilde{N}_m^+ = -\frac{1}{2}N_m/N_{m+1}$, $\tilde{N}_m^- = mN_m/N_{m+1}$, $\tilde{M}_m^+ = \tilde{N}_m^+\tilde{N}_{m+1}^+$, $\tilde{M}_m^0 = \tilde{N}_m^+\tilde{N}_{m+1}^- + \tilde{N}_m^-\tilde{N}_{m-1}^+$, $\tilde{M}_m^- = \tilde{N}_m^-\tilde{N}_{m-1}^-$. Here, note that the following derivations of both $u_m(x)$ and $v_n(y)$ with respect to variables $x$ and $y$ are used to derive Eq.(9), respectively.

$$\frac{du_m(x)}{dx} = \tilde{N}_m^+ u_{m+1}(x) + \tilde{N}_m^- u_{m-1}(x), \qquad (10)$$

$$\frac{dv_n(y)}{dy} = \tilde{N}_n^+ v_{n+1}(y) + \tilde{N}_n^- v_{n-1}(y), \qquad (11)$$

$$\frac{d^2 u_m(x)}{dx^2} = \tilde{M}_m^+ u_{m+2}(x) + \tilde{M}_m^0 u_m(x) + \tilde{M}_m^- u_{m-2}(x), \qquad (12)$$

$$\frac{d^2 v_n(y)}{dy^2} = \tilde{M}_n^+ v_{n+2}(y) + \tilde{M}_n^0 v_n(y) + \tilde{M}_n^- v_{n-2}(y). \qquad (13)$$

Clearly, Eq.(9) is an infinite series equation of both orthogonal functions $u_m(x)$ and $v_n(y)$. Thus, considering the orthogonality of the functions $u_m(x)$ and $v_n(y)$, we obtain a set of equations for the coefficients $\Phi_{m,n}$ of the function $u_m(x)v_n(y)$ by truncating $m=0,1,2,..,M$ and $n=0,1,2,..,N$ in Eq.(9),

$$2\Phi_{0,0}\tilde{M}_0^0 + \Phi_{2,0}\tilde{M}_2^- + \Phi_{0,2}\tilde{M}_2^- = -(a_{1,0}\tilde{N}_1^- + b_{0,1}\tilde{N}_1^-),$$

$$\Phi_{1,0}(\tilde{M}_1^0 + \tilde{M}_0^0) + \Phi_{3,0}\tilde{M}_3^- + \Phi_{1,2}\tilde{M}_2^- = -(a_{0,0}\tilde{N}_0^+ + a_{2,0}\tilde{N}_2^- + b_{1,1}\tilde{N}_1^-),$$

$$\Phi_{0,1}(\tilde{M}_0^0 + \tilde{M}_1^0) + \Phi_{2,1}\tilde{M}_2^- + \Phi_{0,3}\tilde{M}_3^- = -(a_{1,1}\tilde{N}_1^- + b_{0,0}\tilde{N}_0^+ + b_{0,2}\tilde{N}_2^-),$$

$$2\Phi_{1,1}\tilde{M}_1^0 + \Phi_{3,1}\tilde{M}_3^- + \Phi_{1,3}\tilde{M}_3^- = -(a_{0,1}\tilde{N}_0^+ + a_{2,1}\tilde{N}_2^- + b_{1,0}\tilde{N}_0^+ + b_{1,2}\tilde{N}_2^-),$$

$$\Phi_{0,0}\tilde{M}_0^+ + \Phi_{2,0}(\tilde{M}_2^0 + \tilde{M}_0^0) + \Phi_{4,0}\tilde{M}_4^- + \Phi_{2,2}\tilde{M}_2^- = -(a_{1,0}\tilde{N}_1^+ + a_{3,0}\tilde{N}_3^- + b_{2,1}\tilde{N}_1^-),$$

$$\Phi_{0,1}\tilde{M}_0^+ + \Phi_{2,1}(\tilde{M}_2^0 + \tilde{M}_1^0) + \Phi_{4,1}\tilde{M}_4^- + \Phi_{2,3}\tilde{M}_3^- = -(a_{1,1}\tilde{N}_1^+ + a_{3,1}\tilde{N}_3^- + b_{2,0}\tilde{N}_0^+ + b_{2,2}\tilde{N}_2^-),$$

$$\Phi_{0,0}\tilde{M}_0^+ + \Phi_{0,2}(\tilde{M}_2^0 + \tilde{M}_0^0) + \Phi_{2,2}\tilde{M}_2^- + \Phi_{0,4}\tilde{M}_4^- = -(a_{1,2}\tilde{N}_1^- + b_{0,1}\tilde{N}_1^+ + b_{0,3}\tilde{N}_3^-),$$

…

$$\begin{aligned}\Phi_{m-2,n}\tilde{M}_{m-2}^+ + \Phi_{m,n}(\tilde{M}_m^0 + \tilde{M}_n^0) + \Phi_{m+2,n}\tilde{M}_{m+2}^- + \Phi_{m,n-2}\tilde{M}_{n-2}^+ + \Phi_{m,n+2}\tilde{M}_{n+2}^- \\ = -(a_{m-1,n}\tilde{N}_{m-1}^+ + a_{m+1,n}\tilde{N}_{m+1}^- + b_{m,n-1}\tilde{N}_{n-1}^+ + b_{m,n+1}\tilde{N}_{n+1}^-)\end{aligned}. \qquad (14)$$

From Eq.(14), the unknown coefficients $\Phi_{m,n}$ of the perturbation potential $\Phi(x,y)$ can be expressed in terms of the coefficients $a_{m,n}$ and $b_{m,n}$ of the transformation electrical field $E_k^*(x,y)$. If the coefficients $\Phi_{m,n}$ is truncated by the $N$-th order (i.e. $m,n=0,1,2,..,N$), there are $(1+N)^2$ closed equations in Eq.(14) for solving the coefficients $\Phi_{m,n}$. The general solutions of coefficients $\Phi_{m,n}$ can be written as follows,

$$\Phi_{m,n} = \sum_{i,j=0}^{N+1} (g^a_{m,n,i,j} a_{i,j} + g^b_{m,n,i,j} b_{i,j}), \tag{15}$$

where $g^a_{m,n,i,j}$ and $g^b_{m,n,i,j}$ are the functions of the coefficients $N_m$.

For example, the first order approximation solutions of perturbation potential coefficients $\Phi_{m,n}$ (i.e. $m,n = 0,1$) can be derived from the first four equations of Eq.(14) by omitting the higher $\Phi_{m,n}$ (i.e. $m,n > 1$). The non-zero coefficients $g^a_{m,n,i,j}$ and $g^b_{m,n,i,j}$ of the first order approximation solutions $\Phi_{m,n}$ for $N=1$ are given as follows,

$$g^a_{0,0,1,0} = g^b_{0,0,0,1} = -\tilde{N}_1^- / (2\tilde{M}_0^0), \quad g^a_{1,0,0,0} = g^b_{0,1,0,0} = -\tilde{N}_0^+ / (\tilde{M}_0^0 + \tilde{M}_1^0),$$

$$g^a_{1,0,2,0} = -\tilde{N}_0^- / (\tilde{M}_0^0 + \tilde{M}_1^0), \quad g^a_{0,1,1,1} = g^b_{1,0,1,1} = -\tilde{N}_1^- / (\tilde{M}_0^0 + \tilde{M}_1^0),$$

$$g^b_{0,1,0,2} = -\tilde{N}_2^- / (\tilde{M}_0^0 + \tilde{M}_1^0), \quad g^a_{1,1,0,1} = g^b_{1,1,1,0} = -\tilde{N}_0^+ / (2\tilde{M}_1^0), \quad g^a_{1,1,2,1} = g^b_{1,1,1,2} = -\tilde{N}_2^- / (2\tilde{M}_1^0).$$

Here, we note that the higher order approximation solutions of perturbation potential $\Phi(x,y)$ can be easily derived from Eq.(14) by the Mathematica Software.

### 2.3. Solutions of transformation electric fields

To solve the transformation electrical field, substituting Eqs.(15), (7) and (8) into Eq.(3), we obtain the transformation field equations in inclusion regions,

$$e^h E_x^*(x,y) = (e^h - e^i) E_x^0 - (e^h - e^i) \sum_{m,n=0}^{\infty} \sum_{i,j}^{N+1} (g^a_{m,n,i,j} a_{i,j} + g^b_{m,n,i,j} b_{i,j})[\tilde{N}_m^+ u_{m+1}(x) + \tilde{N}_m^- u_{m-1}(x)] v_n(y), \tag{16}$$

$$e^h E_y^*(x,y) = (e^h - e^i) E_y^0 - (e^h - e^i) \sum_{m,n=0}^{\infty} \sum_{i,j}^{N+1} (g^a_{m,n,i,j} a_{i,j} + g^b_{m,n,i,j} b_{i,j})[\tilde{N}_n^+ v_{n+1}(y) + \tilde{N}_n^- v_{n-1}(y)] u_m(x), \tag{17}$$

where $a_{i,j} = \int_{\Omega_i} u_i(x) v_j(y) E_x^*(x,y) dxdy$, $b_{i,j} = \int_{\Omega_i} u_i(x) v_j(y) E_y^*(x,y) dxdy$. It is clear that Eqs.(16) and (17) are equations of transformation electrical fields.

To obtain the solutions of the transformation fields from above equations, the transformation field $E_k^*(x,y)$ can be given as a series, such as a power series,[20,21,31]

$$E_k^*(x,y) = \sum_{i,j=0}^{\infty} C_k^{ij} \left(\frac{x}{L}\right)^i \left(\frac{y}{L}\right)^j, \tag{18}$$

where $k = x, y$. $C_k^{ij}$ is an unknown coefficient. $L$ is the characteristic length of two-dimensional inclusion. Substituting Eq.(18) into Eqs. (16) and (17), and multiplying both sides of the resulting equations by a term $(\frac{x}{L})^k (\frac{y}{L})^l$, and then integrating these equations over the inclusion region $\Omega_i$, we get a set of linear algebra equations about the unknown coefficient $C_k^{ij}$,

$$e^h \sum_{i,j=0}^{\infty} C_x^{ij} A^{i+k,j+l} + (e^h - e^i) \sum_{i,j=0}^{\infty} C_x^{ij} \tilde{H}_a^{i,j,k,l} + (e^h - e^i) \sum_{i,j=0}^{\infty} C_y^{ij} \tilde{H}_b^{i,j,k,l} = (e^h - e^i) E_x^0 A^{k,l}, \quad (19)$$

$$e^h \sum_{i,j=0}^{\infty} C_y^{ij} A^{i+k,j+l} + (e^h - e^i) \sum_{i,j=0}^{\infty} C_x^{ij} \tilde{F}_a^{i,j,k,l} + (e^h - e^i) \sum_{i,j=0}^{\infty} C_y^{ij} \tilde{F}_b^{i,j,k,l} = (e^h - e^i) E_y^0 A^{k,l}, \quad (20)$$

where $\tilde{H}_a^{i,j,k,l} = \sum_{s,t=0}^{N+1} H_a^{k,l}(s,t) G^{i,j}(s,t)$, $\tilde{H}_b^{i,j,k,l} = \sum_{s,t=0}^{N+1} H_b^{k,l}(s,t) G^{i,j}(s,t)$

$$H_a^{k,l}(s,t) = \sum_{m,n=0}^{\infty} g_{m,n,s,t}^a D_{x,m,n}^{k,l}, \quad H_b^{k,l}(s,t) = \sum_{m,n=0}^{\infty} g_{m,n,s,t}^b D_{x,m,n}^{k,l},$$

$$\tilde{F}_a^{i,j,k,l} = \sum_{s,t=0}^{N+1} G^{i,j}(s,t) F_a^{k,l}(s,t), \quad \tilde{F}_b^{i,j,k,l} = \sum_{s,t=0}^{N+1} G^{i,j}(s,t) F_b^{k,l}(s,t)$$

$$F_a^{k,l}(s,t) = \sum_{m,n=0}^{\infty} g_{m,n,s,t}^a D_{y,m,n}^{k,l}, \quad F_b^{k,l}(s,t) = \sum_{m,n=0}^{\infty} g_{m,n,s,t}^b D_{y,m,n}^{k,l}$$

$$D_{x,m,n}^{k,l} = \tilde{N}_m^+ G^{k,l}(m+1,n) + \tilde{N}_m^- G^{k,l}(m-1,n), \quad D_{y,m,n}^{k,l} = \tilde{N}_n^+ G^{k,l}(m,n+1) + \tilde{N}_n^- G^{k,l}(m,n-1)$$

$G^{i,j}(s,t) = \int_{\Omega_i} (\frac{x}{L})^i (\frac{y}{L})^j u_s(x) v_t(y) dxdy$, $A^{i,j} = \int_{\Omega_i} (\frac{x}{L})^i (\frac{y}{L})^j dxdy$. Here, note that the following equations are used to derive Eq.(19) and (20),

$$a_{m,n} = \sum_{i,j=0}^{\infty} C_x^{ij} G^{i,j}(m,n), \quad (21)$$

$$b_{m,n} = \sum_{i,j=0}^{\infty} C_y^{ij} G^{i,j}(m,n). \quad (22)$$

Then, the unknown coefficients $C_k^{ij}$ can be solved with the closed Eqs.(19) and (20) if the number of Eqs.(19) and (20), multiplied by the term $(\frac{x}{L})^k (\frac{y}{L})^l$ ($k,l = 0,1,2,\mathbf{K},S$, and $S$ notes the order of the power-law for the closed equations), equals to the number of the unknown coefficient $C_k^{ij}$. From above derivations, it is clear that there is not any limitation for the inclusion shapes. The effects of the complex inclusion geometry on the transformation field can be expressed by the integrations of Hermite polynomial over the inclusion regions. Thus, our formulas of Eqs.(19) and (20) can be used to investigate the composites having the inclusions of arbitrary shapes.

## 3. Effective dielectric properties

For the constitutive relations of linear composites, the effective dielectric constant $e_{ij}^e$ of two-dimensional composites can be defined as the following form,

$$\overline{D}_k = e_{kx}^e \overline{E}_x + e_{ky}^e \overline{E}_y, \tag{23}$$

where $k = x, y$. $\overline{A} = \frac{1}{V}\int_\Omega A dV$ notes the spatial average of quantity $A$ over the whole composite region $\Omega$ with the volume $V$. $E_k$ is the electrical field distribution in the composite system. Furthermore, for the isotropic matrix and inclusion materials, we have the following formula,

$$\frac{1}{V}\int_\Omega (D_k - e^h E_k)dV = \frac{1}{V}\int_{\Omega_i}(D_k^i - e^h E_k)dV = \overline{D}_k - e^h \overline{E}_k . \tag{24}$$

Substituting Eq.(23) and (1) into Eq.(24), we have,

$$e_{kx}^e \overline{E}_x + e_{ky}^e \overline{E}_y - e^h \overline{E}_k = \frac{1}{V}\int_{\Omega_i}(e^i - e^h)E_k dV = f\frac{1}{V_{\Omega_i}}\int_{\Omega_i}(e^i - e^h)E_k dV, \tag{25}$$

where $f = V_{\Omega_i}/V$ is the volume fraction of inclusions, and $V_{\Omega_i}$ is the volume of inclusions. Eq.(25) shows that the effective responses can be calculated by the volume averages of the electrical field over the inclusion regions.

Moreover, we can regard the external applied field $E_k^0$ as the average field $\overline{E}_k$. From Eq.(25), for $k = x$, $\overline{E}_x = E_x^0$ and $\overline{E}_y = 0$ (i.e. the external applied electric field $E_x^0$ along $x$-direction), the following effective dielectric constant formula is given,

$$e_{xx}^e = e^h + f\frac{1}{E_x^0 V_{\Omega_i}}\int_{\Omega_i}(e^i - e^h)E_x dV . \tag{26}$$

In addition, for $k = x$, $\bar{E}_y = E_y^0$ and $\bar{E}_x = 0$, we have,

$$e_{xy}^e = f \frac{1}{E_y^0 V_{\Omega_i}} \int_{\Omega_i} (e^i - e^h) E_x dV, \qquad (27)$$

Similarly, for $k = y$ in Eq.(25), we get,

$$e_{yy}^e = e^h + f \frac{1}{E_y^0 V_{\Omega_i}} \int_{\Omega_i} (e^i - e^h) E_y dV, \qquad (28)$$

$$e_{yx}^e = f \frac{1}{E_x^0 V_{\Omega_i}} \int_{\Omega_i} (e^i - e^h) E_y dV. \qquad (29)$$

Here, note that both effective dielectric constants $e_{yx}^e$ and $e_{xy}^e$ are symmetric for usual composites. For the composites containing both isotropic matrix and inclusions, the effective dielectric constants are not always isotropic, and depend on the structures and distributions of inclusions. Clearly, the effects of the inclusion geometry and distribution on the effective responses are shown in Eq.(26), (27) and (28). Moreover, based on the Eq.(3), the electrical fields $E_k^i(x, y)$ of the inclusion regions are expressed by the transformation electrical fields,

$$E_k^i(x, y) = E_k^0 + E_k^p(x, y) = \frac{e^h}{e^h - e^i} E_k^*(x, y), \qquad (30)$$

where $k = x, y$. Thus, substituting Eq.(30) into Eqs.(26), (27) and (28), the effective dielectric constants can be calculated. Next, these formulas are applied to calculate the effective dielectric constants of isotropic composites.

## 4. Numerical discussion

To verify the validity of our method, as an example of two-dimensional problem, the effective dielectric responses of the isotropic cylindrical composite, where a long cylindrical inclusion with radius $a$ (i.e. radius $a$ within the $x-y$ plane) is embedded in an infinite matrix, are estimated. In our numerical calculation, the isotropic dielectric constants of matrix and inclusion are $e^h = 25$ and $e^i = 15 + n$, respectively, where $n$ is the parameter of the inclusion dielectric constants. The dimensionless radius $a$ of the cylinder inclusion is $a=2$, and the volume fraction of inclusions is $f=0.05$. The origin of Cartesian coordinates $(x, y)$ is located at the center of cross section of the cylindrical inclusion. Under external applied electrical field $E_x^0$ along the $x$-direction (or $E_y^0$ along the $y$-direction), our method is applied to calculate the effective responses. The results of the method are compared with the exact dilute solutions of cylindrical composites [35].

In Table 1, the effective dielectric constants $e_{xx}^e/e^h$ are calculated by the method on the basis of the zero-th order and the first order approximations of the perturbation potential $\Phi(x, y)$ given in Eq.(15), respectively. Good agreements are obtained by comparing with the exact dilute solutions of the cylindrical composites [35]. It is shown that the results of first order approximation are more accurate than that of the zero-th order approximation to tend the exact dilute solutions. The results also indicate that the high order solutions should give excellent performance for estimating the effective properties of the composites. Here, note that, in our calculation, the effective dielectric constant $e_{yy}^e$ equals to $e_{xx}^e$ due to the symmetry of the cross section of the cylinder in the $x-y$ plane, and both effective dielectric constant $e_{xy}^e$ and $e_{yx}^e$ are zero. Hence, the numerical results demonstrate that the extended Eshelby's method is valid to cope with the effective response problem of composites having arbitrary inclusion geometry since the transformation field method does not has any limitation for the inclusion shapes.

**Table 1.** Effective dielectric constant ratio $R$ (i.e. $e_{xx}^e/e^h$) estimated by the 0-th order and the first order perturbation potentials for different factor $n$ of inclusion dielectric constant, where the dielectric constants of matrix and cylindrical inclusion are $e^h=25$, $e^i=15+n$ with the inclusion dielectric parameter $n$, respectively. Here the inclusion volume fraction is $f=0.05$. In this table, "exact" notes the exact solution of cylindrical dielectric composite in the dilute limit.

| $n$ | 0-th order | first order | exact |
|---|---|---|---|
| 1 | 0.98200 | 0.98148 | 0.97805 |
| 2 | 0.98400 | 0.98360 | 0.98095 |
| 3 | 0.98600 | 0.98569 | 0.98372 |
| 4 | 0.98800 | 0.98778 | 0.98636 |
| 5 | 0.99000 | 0.98985 | 0.98889 |
| 6 | 0.99200 | 0.99190 | 0.99130 |
| 7 | 0.99400 | 0.99395 | 0.99362 |
| 8 | 0.99600 | 0.99598 | 0.99583 |
| 9 | 0.99800 | 0.99799 | 0.99796 |
| 10 | 1.00000 | 1.00000 | 1.00000 |
| 11 | 1.00200 | 1.00199 | 1.00196 |
| 12 | 1.00400 | 1.00398 | 1.00385 |
| 13 | 1.00600 | 1.00595 | 1.00566 |
| 14 | 1.00800 | 1.00791 | 1.00741 |
| 15 | 1.01000 | 1.00986 | 1.00909 |
| 16 | 1.01200 | 1.01180 | 1.01071 |
| 17 | 1.01400 | 1.01373 | 1.01228 |
| 18 | 1.01600 | 1.01565 | 1.01379 |
| 19 | 1.01800 | 1.01756 | 1.01525 |
| 20 | 1.02000 | 1.01946 | 1.01667 |

## 5. Conclusions

Eshelby's transformation field method is developed to solve the open boundary problem of composites having arbitrary geometric inclusions. The transformation fields are introduced in the composite system to deal with the complex interface conditions between the inclusion and the matrix. The transformation electric fields and the perturbation potential are expressed by Hermite polynomials for cope with the open boundary conditions. Moreover, a set of algebraic equations is given for solving the transformation field on the basis of the transformation field equations. As an example, the method is used to estimate the effective dielectric responses of the two-phase isotropic composites having open boundary conditions, and the formulas of the effective dielectric constants are expressed by the transformation fields. Furthermore, the validity of the method is verified by comparing the effective responses calculated by the zero-the order and the first order approximation solutions of perturbation potentials with the exact solutions of dilute limit. The numerical results show that the method is valid to deal with the open boundary problem of the composites having complex geometric inclusions. As a brief conclusion, it is noted that the higher order solutions of the perturbation potentials given by Hermite polynomials will obtain more accurate results. In addition, the method can be used to study the open boundary problem of the anisotropic composites, such as dielectric, elastic, thermal, piezoelectric composites, etc.

**Acknowledgements**

This work was supported by NSFC (Grant No. 42276178).